\documentstyle[twocolumn,pra, aps, epsf]{revtex}  
\begin{document}

\title {Sensitive Absolute Gravity Gradiometry Using Atom Interferometry}

\draft

\author{J. M. McGuirk, G. T. Foster, J. B. Fixler, M.J. Snadden, and M. A. Kasevich}
\address{Physics Department, Yale University, New Haven  CT, 06520}
\date{\today}
\maketitle

\begin{abstract}
We report the demonstration of a sensitive absolute gravity gradiometer based on light-pulse atom interference techniques. The
gradiometer consists of two absolute accelerometers operated in a
differential mode.  We report a differential acceleration sensitivity of $4\times 10^{-9}$ g/Hz$^{1/2}$ and an inferred differential acceleration accuracy of less than 10$^{-9}$ g.  This corresponds to a gravity gradient sensitivity of 4 E/Hz$^{1/2}$ (1 E = 10$^{-9}$ s$^{-2}$) and an accuracy of better than 1 E for a 10 m separation between accelerometers.  We demonstrate that the
instrument can be used to detect nearby masses in a vibrationally
noisy environment and characterize instrument sensitivity to
spurious acceleration and rotation noise.

\end{abstract}


\section{Introduction}

Precision gravimetry is scientifically and technologically relevant.  For example, theories which predict violations of Einstein's general relativity are manifested in gravitational phenomenon such as composition dependent gravitational forces, time variation of the gravitational constant G, or breakdown of the 1/r$^2$ law \cite{dam94}.  Some theories predict a fifth force originating from a spin-gravity coupling \cite{ven92}. The gravitational constant itself is only known to a few parts in 10$^{-4}$ \cite{moo00}.  Technological applications lie in the fields of navigation, geodesy, underground structure detection, and oil and mineral exploration.

A technical problem associated with the characterization of gravitational forces is rooted in the equivalence principle: in
principle it is not possible to distinguish gravitationally-induced accelerations from accelerations of the reference frame of measurement.  In other words, gravimetry is fundamentally challenged by the inability to distinguish platform vibrations from gravitational accelerations.  It is well known that this difficulty can be overcome with gradiometric measurements in which two simultaneous, spatially separated acceleration measurements are made with respect to a common reference platform. The difference between these acceleration
measurements suppresses platform vibration noise as a common mode,
but preserves gravitationally-induced differential accelerations
(gravity gradients).  This non-local measurement of the curvature
of the gravitational potential circumvents the equivalence principle and allows for characterization of gravitational anomalies in vibrationally noisy environments.  For this reason gravity gradiometers are of technological interest.

In this article we present a sensitive absolute gravity gradiometer based on atom interference methods. We demonstrate a differential acceleration sensitivity of $4\times 10^{-9}$ g/Hz$^{1/2}$ (corresponding to an inferred performance of $2.8\times 10^{-9}$
g/Hz$^{1/2}$ per accelerometer). We characterize instrument immunity to rotational and vibrational noise through laboratory shake-tests. We measure the gravitational gradient induced by nearby mass distributions and characterize instrument accuracy. In previous work we made a proof-of-principle measurement of the gravitational gradient of the Earth using a substantially less sensitive instrument \cite{sna98}.

The central idea behind our atom interferometer gravity gradiometer is to make acceleration measurements on two vertically separated laser cooled ensembles of cesium (Cs) atoms in free-fall using a pair of vertically propagating laser beams.  The propagation axes of these laser beams are aligned to pass through both ensembles.  The light-pulse atom interference method is used to measure the acceleration of each ensemble with respect to a reference frame defined by the phase fronts of the interrogating optical fields. The difference between the measured acceleration of each atom ensemble, divided by their separation, is a measure of the in-line component of the gravity-gradient tensor T$_{ij}$. (This tensor characterizes the gravitational field inhomogeneity induced by non-uniform mass distributions.) Accelerations of the common reference frame -- defined by the optical field phase fronts -- are rejected as a common-mode in the difference.

Each acceleration measurement is accomplished by driving an
optical pulse $\pi/2-\pi-\pi/2$ atom interferometer sequence
between the 6S$_{1/2}$, F = 3 and F = 4 ground-state hyperfine
levels of atomic Cs \cite{ber97}.  This method has recently been
used by Chu and coworkers  to measure g at the part per billion
level, with a sensitivity of $ 2\times 10^{-8}$ g/Hz$^{1/2}$
\cite{peters99}.

The interferometer theory is described in detail elsewhere
\cite{ber97,kas92}.  Here we review the essential results.  Each
interferometer pulse couples two ground-state hyperfine levels via
a two photon, velocity-sensitive Raman coupling \cite{kas91b}. In
the short, intense pulse limit, the optical phase associated with
the Raman coupling is imprinted on the atomic center-of-mass
wavepackets.  This phase locates these wavepackets with respect to
the reference frame defined by the optical field.  The following
rules, derived from solutions of the Schr\"{o}dinger equation for a
two-level atom in the rotating wave approximation, govern this
imprinting process: $|3 \rangle \rightarrow e^{i \phi(t)} |4
\rangle$ and $|4 \rangle \rightarrow e^{-i \phi(t)} |3\rangle$,
where $\phi(t)$ is the phase of the driving field at the mean
position $\bf{x}$ of the wavepacket at the time $t$ of the
interferometer pulse, and states $|3 \rangle$ and $|4\rangle$ are
the two resonantly coupled hyperfine states.  Explicitly, $\phi(t)
= {\bf k_{eff}} \cdot {\bf x}(t) + \phi_0(t).$  Here the Raman
propagation vector ${\bf k_{eff}}$ is defined as ${\bf k_1}-{\bf
k_2}$; ${\bf k_1}$ and ${\bf k_2}$ are the wavevectors for two
Raman lasers.

The probability following the three pulse interferometer sequence
for the atoms to be found in $|4\rangle$ (if they initially are
prepared in $|3\rangle$) can be evaluated from straightforward
application of the above rules and is given by P$(|4\rangle)$ = [1
- cos $\Delta\phi$]/2.  The phase shift $\Delta\phi \equiv
\phi(t_1) - 2\phi(t_2)+\phi(t_3)$, where t$_i$ is the time of the
i$^{th}$ pulse.  Evaluating $\phi(t_i)$ as in \cite{ber97} gives
$\Delta\phi = {\bf k_{eff} \cdot a}$T$^2 + \Delta\phi_0$.   T is
the interrogation time between interferometer pulses, {\bf a} is
the mean acceleration experienced by the atoms with respect to the
optical fields, and $\Delta \phi_0 = \phi_0(t_1) - 2\phi_0(t_2)+\phi_0(t_3)$.  For vibrationally quiet environments,
$\phi_0(t_1) = \phi_0(t_2) = \phi_0(t_3)$.  In this case, $\Delta
\phi_0 = 0$, and measurement of the transition probability
following the three pulse sequence determines ${\bf a}$.  In
noisier environments, $\Delta \phi_0$ is no longer zero, and this
spurious shift contaminates an individual acceleration measurement.

The gradient signal is derived from two simultaneous, spatially-separated, acceleration measurements which are made with respect to the same set of Raman laser fields.  This is accomplished by simultaneously measuring the fraction of atoms excited by the pulse sequence at both spatial locations. The differential acceleration is given by the differential phase shift, $\Delta\phi = {\bf k_{eff}} \cdot \Delta{\bf g}$T$^2$, between the upper and lower atom ensembles, where $\Delta{\bf g}$ is the difference in the mean gravitational acceleration at the two accelerometers.  In the framework of this analysis, low frequency platform accelerations introduce common accelerations $\delta {\bf a}$ which cancel in the differencing procedure. High frequency vibrational noise, which shows up in the $\Delta \phi_0$ terms of each accelerometer, also cancels.

Our approach differs from that used in other instruments.  First,
it uses atoms as proof masses rather than macroscopic objects.
This eliminates variability from device-to-device and provides
insensitivity to many environmental perturbations, {\it e.g.}
temperature fluctuations and magnetic fields.  The gradiometer
references its calibration to the wavelength of the measurement
laser, which is locked to an atomic spectral line, providing
absolute accuracy and long term stability. Because the sensitive
axis is defined by the Raman propagation vector, and the
acceleration measurements are referenced only to one retroreflector, the two accelerometers may be placed far apart without sacrificing common-mode vibration rejection.  Increasing separation between accelerometers linearly increases the sensitivity to gravity gradients and provides insensitivity to near-field perturbations.

The gradiometer presented here performs favorably when compared with existing gravity gradient sensors.  The state-of-the-art mobile gradiometer is a device that uses mechanical mass-spring accelerometers on a rotating, gimballed fixture (Lockheed Martin UGM) \cite{jek93}.  This device has recently been used to perform airborne surveys of geophysical mass anomalies \cite{bhp} and to characterize the gravitational anomalies from man-made underground structures \cite{rom}.  Higher sensitivity laboratory devices based on superconducting transducers have achieved significantly better sensitivities ($< 0.1$ E/Hz$^{1/2}$) \cite{moo93,van94}.  However, these devices suffer from tare effects in the superconductors \cite{goo99} and their reliance on cryogens.  More recently, the differential acceleration of two falling corner cubes has been measured using a Michelson interferometer configuration \cite{bro00}.  However, the demonstrated sensitivity of this device of 450 E/Hz$^{1/2}$ is not competitive with the Lockheed Martin UGM, and it remains to be seen if this device can demonstrate the common-mode noise rejection required for mobile applications.

This remainder of this paper is organized as follows.  Section II details the apparatus. Section III contains results characterizing the instrument's performance. Section IV draws comparisons with related  atom interference methods. Finally, future work is detailed in Section V.

\section{Apparatus}

\subsection{Apparatus Overview}
The apparatus is similar to that described in Ref. \cite{sna98},
but several changes have been made in order to achieve the current
sensitivity.  The apparatus consists of two laser cooled and trapped
sources of Cs atoms. The atoms are launched on ballistic
trajectories and prepared in a particular internal state with
optical and microwave techniques before undergoing the
interferometer sequence. Following the interferometer sequence,
atoms are detected using a normalizing detection method. Each of
these stages is described below.  In addition, we detail the
operation of an actively controlled vibration isolation system
which is used to validate our data acquisition methodology.

\subsection{Laser Cooled Atomic Sources}
Each laser cooled atomic source consists of an ultra-high vacuum
system with a Cs source which maintains a low Cs vapor pressure
for a magneto-optical trap (MOT), along with the required laser
beams and magnetic field configuration to form the MOTs.  The
lasers are delivered to the vacuum chambers via optical fibers
from a common high-power, frequency stable laser system.  The
laser beams are configured in the six beam [1,1,1] geometry (3
mutually orthogonal pairs) which allows clear access for the Raman
interferometer beams along the the vertical axis.

In order to obtain good interferometer signal-to-noise ratios
(SNRs), it is critical to load atoms quickly into the MOTs in a
way which minimizes atom number fluctuations during the loading
process.  A grating stabilized diode laser (New Focus Vortex)
replaces the previously used distributed Bragg reflector (DBR)
laser as the master laser for the trapping system. The extended
cavity of the Vortex laser gives it an intrinsic linewidth
approximately ten times smaller than the DBR laser.  The Vortex
laser is locked to a Cs transition via standard saturation
spectroscopy techniques in a Cs cell.  The locked laser has a
stability of 1 kHz/Hz$^{1/2}$ and an instantaneous linewidth of
300 kHz.  The more stable master laser reduces the shot-to-shot
rms number fluctuations of the MOTs by a factor of five to a SNR
of 200:1.  In addition, frequency-induced noise during the
detection of the atoms following the interferometer is
substantially suppressed with the use of this laser system.

The master laser injection locks two amplifier lasers which then
seed two 500 mW tapered amplifier lasers (SDL 8630E).  All of the
injection locks transfer the frequency stability of the master
laser, without addition of extra frequency or intensity noise.
Each tapered amplifier laser is split into six equal outputs to
provide the twelve trapping beams for the two MOTs. The splitting
is accomplished using robust and compact free-space
fiber optic beamsplitters (Optics for Research FiberBench).  The
stability of the fiber mounts maintains the splitting ratio to
greater than 1$\%$ over many months with no adjustment.  The
fibers are polarization maintaining with an extinction ratio of
greater than 30 dB.  Polarization stability is further maintained
with the use of clean-up polarizing optics at the output of the
fibers (converting polarization drift in the fiber to negligible
intensity fluctuations). The intensity and polarization stability
of the MOT trapping beams contribute greatly to the reduction of
the MOT number fluctuations.  The trapping beams from the twelve
fibers propagate uncollimated to the two MOTs.  By not collimating
the beams, we are able to circumvent window apertures limiting the
beam size.  Larger beam waists in the loading region lead to
higher atom loading rates. The approximate beam waist at the
trapping position is 2.5 cm (1/e$^2$), and the intensity is about
1.2 I$_{sat}$ per beam (I$_{sat}$ = 1.1 mW/cm$^2$ for the Cs
cooling transition). In this configuration, each MOT loads
approximately $2\times10^8$ atoms in 1 s.

After loading the MOTs from a thermal vapor for 1 s, the cold
atoms are launched in ballistic, atomic fountain trajectories. The
launch is accomplished as follows.  First, the detuning of the MOT
beams is switched from -10 MHz to -20 MHz from the cooling
transition. Next, the trapping quadrupole magnetic fields are
turned off, and the atoms are held in the -20 MHz detuned optical
molasses for 30 ms while the eddy currents from the field
switching damp out (the vacuum chambers are aluminum). For each
MOT, following this holding period, the frequency of the upper
three molasses beams is ramped down by 1 MHz over 5 ms [by
applying the appropriate RF frequency shifts to an acousto-optic
modulator (AOM)], while the frequency of the lower three beams is
ramped up by an equal amount.  This frequency ramp smoothly
transfers the atoms to an optical molasses moving vertically at
~1.5 m/s.  After this ramp, the mean frequency of the trapping
beams is ramped down to -40 MHz detuned (still in a moving
molasses) over 5 ms, and then the intensity is ramped down to half
intensity in 1 ms, held for 0.5 ms, and ramped completely off in
an additional 0.5 ms.  These ramps cool the atoms to  ~2.3 $\mu$K
(measured with stimulated Raman velocimetry).  The frequency and
intensity ramps are accomplished using digitally synthesized
waveforms from Hewlett Packard HP8770A arbitrary waveform
generators (AWGs).

Following this launching and cooling phase, the cold atoms move in
a 320 ms, 12 cm high fountain during which they are prepared in a
special state and then interrogated by the interferometer
sequence.

\subsection{State Preparation}
Before the atom interference pulse sequence, a state selection
sequence prepares the atoms in the magnetically insensitive m$_f =
0$ sublevel and velocity selects an atomic ensemble with a
velocity spread which is matched to the velocity selectivity of
the Raman pulse sequence. This state selection, which is
accomplished with a sequence of microwave and optical pulses, is
important for obtaining high fringe contrast.  The details
associated with this state selection are discussed below.

Following their launch, atoms are initially distributed nearly
evenly among the magnetic sublevels of the F = 4 ground state.
Three orthogonal pairs of magnetic field coils, roughly in a
Helmholtz configuration, zero the Earth's magnetic field and apply
a vertical bias field of $\sim$ 100 mG.  This bias allows
selective addressing of individual F = 3 to F = 4 ground state
hyperfine transitions with a 9.2 GHz microwave field (delivered to
the atoms through a Narda 640 X-band gain horn).  First, a
microwave composite $\pi$ pulse transfers atoms from the F = 4,
m$_f$ = 0 to the F = 3, m$_f$ = 0 sublevel. (Composite pulse
sequences accomplish the population transfer associated with ideal
$\pi$ in an experimentally robust way, as described in the
subsection below.) Next, a near-resonant pulse from the upper
trapping beams tuned slightly above the 6S$_{1/2}$, F =4
$\rightarrow$ 6P$_{3/2}$, F$^\prime$ = 5 transition clears atoms
in the remaining F = 4 sublevels (via the scattering force).
Another composite microwave $\pi$ pulse then returns F = 3, m$_f$
= 0 atoms to F = 4, m$_f$ = 0. An optical velocity selective
composite Raman $\pi$ pulse is now applied which transfers F = 4,
m$_f$ = 0 atoms within the velocity range encompassed by the Raman
pulse envelope to F = 3, m$_f$ = 0. Finally, a second, near
resonant, blue detuned pulse clears away the remaining F = 4
atoms. At this point, the remaining (state prepared) atoms are
ready for use in the interferometer.

A considerable fraction of atoms are eliminated from the initial
ensemble in this process:  $\sim$ 8/9 from the internal state
selection and another $\sim$ 2/3 in velocity selection, leaving
roughly 4\% for the interferometer.

\subsubsection{Enhanced Optical Pumping}
In order to increase the m$_f$ = 0 population, an enhanced optical
pumping scheme has been implemented.  This method essentially
recycles atoms which are not initially in the m$_f$ = 0 target
state by a cyclic optical pumping sequence.  In practice we have
seen as much as a factor of 3 improvement in usable atoms with
this method.

The details of this scheme are as follows.  First, a composite
microwave $\pi$ pulse is applied to drive atoms from the F = 4,
m$_f$ = 0 to the F = 3, m$_f$ = 0 state as before. Then a
de-pumping beam tuned to the  F = 4 to F$^\prime$ = 4 transition
is applied to optically pump the remaining atoms from F = 4, m$_f
\not=$ 0 to F = 3 (where the prime still represents the 6P$_{3/2}$
excited manifold).  The de-pumping process redistributes the
atomic population, with approximately 1/7 of the remaining atoms
ending up in m$_f$ = 0. The process is then reversed with a
composite microwave $\pi$ pulse transferring F = 3, m$_f$ = 0 to F
= 4, m$_f$ = 0 followed by application of a repumping beam to the
F = 3 to F$^\prime$ = 4 transition.  In principle, this entire
sequence could be repeated many times, resulting in 100\% pumping
efficiency.

In practice, inefficiency of the microwave pulses (described
below), heating due to spontaneous emission in the pumping
sequence, and the availability of only a finite amount of time to
execute the sequence limit the process efficiency.  We realize a
factor of three improvement with just one cycle of the sequence
described in the previous paragraph.

For vapor cell loaded traps, the overall efficiency is also
limited by the presence of the background atomic vapor. In this
case, the de-pumping photons can excite atoms in the background
vapor which then emit light at the repumping frequency. These
rescattered photons then redistribute atoms in the F = 3, m$_f$ =
0 state to other m$_f$ $\not=$ 0 states.  Thus there is a
trade-off between background vapor pressure (which sets the
loading rate) and the overall efficiency of the scheme (which
works best at low vapor pressure).  For our operating parameters,
we typically realize a factor of 1.5 - 2 improvement.

Finally, we note that this method substantially increases
shot-to-shot atom number fluctuations (from 200:1 to 60:1).
However, our normalized detection method is able to effectively
suppress the impact of these fluctuations on the interferometer
signals.

\subsubsection{Composite Pulse Techniques}
The above state preparation methods work best with efficient
coherent population transfer between F = 3 and F = 4 states.  Less
than unit transfer efficiency during a standard $\pi$ pulse
between the ground states can result from an inhomogeneous Rabi
frequency of the microwave or optical pulse seen by the atoms, as
well as by detunings due to the velocity spread of the atoms.  In
our apparatus, the microwave $\pi$ pulses are typically only 80\%
efficient due to inhomogeneous field strengths across the atom
clouds (since we drive the microwave transition with horns located
outside the vacuum chamber). Furthermore, the state selection and
optical pumping require a series of one optical and four microwave
$\pi$ pulses in the two separate chambers.  With the current
system, it is difficult to match the pulse conditions for all
pairs of pulses due to different microwave intensity gradients in
each chamber.

In composite pulse sequences\cite{levitt86}, a standard $\pi$ pulse is replaced with a sequence of pulses with variable area and relative phase.  In our work, we employ a $\pi/2 - \pi_{90^{\circ}} - \pi/2$ pulse sequence in place of a $\pi$ pulse. The subscript ${90^{\circ}}$ indicates that the phase of the center $\pi$ pulse is shifted 90$^{\circ}$ relative to the $\pi/2$ pulses.  The use of this sequence increases the transfer efficiency of a pulse for inhomogeneous distributions of Rabi frequency and detuning across the atomic ensemble.  Employing these pulses for the microwave state preparation pulses increases the transfer efficiency from 80$\%$ with a regular $\pi$ pulse to 95$\%$.  Fig.~\ref{fig1} compares a frequency scan with regular and composite $\pi$ pulses. More advanced interferometer pulse sequences may benefit from the use of composite pulses for the optical pulses but are not employed at the moment (see Section VI.A).

In comparison with the adiabatic rapid passage technique (ARP)
\cite{abragam61}, composite pulses are easier to implement
experimentally, and require less time (or total pulse area) to
achieve efficient transfer.  For example, we find that we need
approximately $5 \pi$ total time (here time is referenced to the
time to drive a $\pi$ pulse at the maximum Rabi frequency) for ARP
to achieve results similar to those achieved with only $2 \pi$
time for the composite sequence.

\subsection{Atom Interferometer}
Following its launch and state preparation each atom ensemble is
subject to the $\pi/2-\pi-\pi/2$ interferometer pulse sequence.
Key experimental details associated with this sequence include the
frequency stability of the lasers used to drive the Raman
transition, optimization of the Raman excitation parameters, and
the physical geometry for beam delivery.

\subsubsection{Raman Lasers}
As the acceleration-induced phase shifts depend critically on the
phase and wavevector associated with the laser beams used to drive
stimulated Raman transitions, it is important to evaluate the
possible contributions of laser frequency noise on the relative
stabilities of the interferometer signals.  We first consider the
constraints on laser frequency stability, then describe the laser
system we use to meet these constraints.

In our excitation geometry, the two ensembles are separated by
$\sim$ 1.4 m.  The Raman fields propagate in an asymmetric way to
these ensembles (see Fig.~\ref{fig2}).  To see this, consider the
propagation paths of the optical fields following the beam
dividing optic which separates the two Raman fields.  The beam of
frequency $\nu_1$ propagates roughly $x_1^1 \approx$ 0.3 m before
it passes through the first ensemble of atoms, while it propagates
roughly $x_1^2 \approx$ 1.7 m before it passes through the second
ensemble.  On the other hand, the beam of frequency $\nu_2$
propagates $x_2^1 \approx$ 3.7 m before it passes through the
first ensemble, while it propagates $x_2^2 \approx$ 2.3 m before
it passes through the second ensemble.

If the frequency of the lasers drifts on a time scale short
compared to the interrogation time T between pulses, this can
cause an asymmetric phase shift to be read into the atomic
coherences due to this path asymmetry.  For example, suppose the
laser frequency jitter is $\delta \nu$, while the differential
path length travelled by the Raman lasers for the two ensembles is
$\ell$  $\equiv$ (effective path to ensemble 1) - (effective path
to ensemble 2) = ($x_2^1-x_1^1$) - ($x_2^2-x_1^2$) $\simeq 2.8$ m.
This leads to a differential phase noise of $\delta \phi_{laser}
\sim (k \ell)(\delta \nu/\nu)$.  For a target interference SNR of
1000:1, $\delta \phi_{laser} \lesssim$ 1 mrad. For our parameters,
this implies $\delta \nu \lesssim$ 20 kHz.

In order to achieve frequency stability at this level, a second
Vortex (external cavity, grating stabilized) laser is used as the
master laser for the Raman laser system.  This laser is locked to
the 6S$_{1/2}$, F = 3 $\rightarrow$ 6P$_{3/2}$, F$^\prime$ = 2
crossover resonance (via modulation transfer spectroscopy) using several AOMs to offset the frequency to obtain the desired detuning.  The lock is maintained through the use of a digital signal processor
(DSP, Spectrum Signal Indy TMS320C32).  The DSP processes the lock
error signal through a highpass and lowpass channel, each
operating at a sampling rate of 25 kHz. The high and low frequency
channels provide feedback to the laser current and to the cavity
piezo element respectively.  Due to the presence of long term
drift in the piezo element a third, very low frequency channel is
provided through a GPIB command to the laser controller.  The measured
stability of the laser is comparable to that of the master
laser used for the optical molasses (and is primarily limited by a
5 kHz resonance of the laser's piezo-electric transducer).

The Vortex laser directly injection locks a 150 mW (SDL 5422) slave diode laser. This laser is then shifted up and down in frequency by 4.756 GHz (160 MHz above half the Cs clock frequency) with a high frequency AOM. The diffracted orders are then used to injection lock two more 150 mW slave laser diodes at a frequency 700 MHz red detuned from the F = 3 $\rightarrow$ F$^\prime$ = 4 and F = 4 $\rightarrow$ F$^\prime$ = 4 transitions, respectively \cite{bou96}.  The frequency noise on the master Raman laser exists on both the slave lasers, but their frequency separation remains fixed at the Cs ground state hyperfine transition. The detuning from the F$^\prime$ = 5 level reduces the effect of spontaneous emission due to off-resonant single photon excitations from each Raman beam.

\subsubsection{Raman Beam Delivery}
The Raman beams are delivered to the vacuum chambers with a polarization maintaining optical fiber in order to increase the pointing stability of the Raman beams as well as to spatially filter them.  Prior to coupling into the fiber, the two Raman beams are double-passed through 80 MHz AOMs.  These AOMs are controlled by another HP8770 AWG which allows dynamic frequency, phase, and intensity tuning of the Raman beams during the interferometer pulse sequence.  During the interferometer, the frequency of the Raman lasers must be chirped in order to maintain a resonance condition with the accelerating (Doppler shifting) atoms.

The two Raman beams are overlapped with orthogonal linear polarizations on a polarizing beamsplitting cube and passed through a Pockel's cell polarization modulator (ConOptics 350-50) after the double-passed AOMs.  The Pockel's cell is used to reverse the effective Raman laser propagation direction, as described below. The Raman beams then are coupled into a polarization maintaining fiber with 75\% efficiency and sent to the gradiometer.  For wavefront quality, after the fiber the Raman beams are collimated with a 1.1 cm focal length aspheric lens in conjunction with a high surface quality 50 cm focal length spherical lens.  This lens combination results in a uniform phase front for the Raman beams. After collimation the Raman beams have a 1.0 cm (1/e) beam waist. All optics in the Raman beams' propagation path after the optical fiber are of high surface figure ($\lambda$/10 or better) in order to preserve the phase front homogeneity of the Raman beams.

After collimation, the Raman beams enter a racetrack geometry in order to obtain counterpropagating beams for the Raman interaction (see Fig.~\ref{fig2}).  The racetrack configuration starts with a polarizing beamsplitter cube that separates the two orthogonally polarized Raman beams.  The two Raman beams then parallel propagate vertically through the vacuum chambers with one beam passing through the axis of the atom ensembles and the other 2 cm off-axis.  After passing through both chambers, a corner cube retroreflector redirects the off-axis Raman beam to counterpropagate with the on-axis Raman beam, resulting in two counterpropagating beams.   The use of the corner cube decreases the tilt sensitivity of the apparatus by keeping the Raman beam propagation axis constant as the cube is subjected to spurious tilts.  In this racetrack, standing waves are eliminated, which maintains overall intensity stability by suppressing etalon effects from the Raman beams.  Also, spontaneous emission is reduced by half compared to using collinear retroreflected beams.

\subsubsection{Raman Propagation Reversal}
A technique to reduce many systematic interferometer phase shifts involves reversing the effective Raman propagation vector ${\bf k_{eff} = k_1-k_2}$.  Because the gravitational phase shift is proportional to ${\bf k_{eff} \cdot g}$, reversing the sign of ${\bf k_{eff}}$ changes the sign of the gravitational phase shift. However, several systematic phase shifts, such as second-order Zeeman shifts from magnetic fields and any residual AC Stark shifts, have no dependence on the Raman wavevector direction. Subtracting the phases obtained from consecutive experimental cycles using two reversed propagation directions gives twice the gravitational phase shift, but removes these systematic shifts. The propagation reversal is accomplished using the Pockel's cell. The Pockel's cell rotates the polarization of both Raman beams by
$90^\circ$ when activated.  This rotation causes the direction the Raman beams take through the racetrack to switch, {\it i.e.} $\bf{k_{eff}} \rightarrow \bf{-k_{eff}}$. Propagation direction can be switched from one shot to the next.

\subsubsection{Raman Beam Parameters}
A theoretical study of the interferometer contrast as a function
of Raman beam waist and detuning was performed.  A finite Raman
beam size gives rise to a spatially inhomogeneous Rabi frequency
across the atom cloud, causing dephasing during the interferometer.  Similarly, the velocity spread of the atoms along the Raman beams causes inhomogeneous broadening due to differential Doppler shifts across the atom ensemble.  Typical Rabi frequencies are around 30 kHz, and the initial width of the thermal spread of the atom ensemble is about 45 kHz. Incorporating the finite beam size and the initial atomic velocity distribution function into our analysis, we find an optimal 35\% contrast for a Raman beam waist of 1.0 cm radius (1/e), and a detuning of $\approx$ 1 GHz from F$^\prime$ = 5 (our chosen operating point).

In addition to the interferometer contrast, systematic AC Stark
shifts were studied.  AC Stark shifts from the Raman pulses
themselves cause spurious phase shifts in the interferometer if
unconstrained.  However, with a two-photon Raman transition, the
AC Stark shift is the difference between the individual AC Stark
shifts from each beam and can be zeroed by adjusting the ratio
between the two Raman beams. The Stark shift is balanced with a
beam intensity ratio of $\sim$ 1.6:1 for the chosen Raman
detuning.  The Stark shift is balanced empirically by inserting
off-resonant Raman pulses within a microwave $\pi/2-\pi-\pi/2$
interferometer and adjusting the Raman beam intensity balance to
zero the optically induced phase shift.  This ratio of 1.6:1 for
the chosen Raman detuning agrees well with theoretical
predictions.

\subsubsection{Interferometer Operation}
The gradiometer is typically operated in its most sensitive
configuration with the interferometer pulses at a maximal spacing
of T = 157.5 ms.  This time is limited by the vacuum chamber size
which constrains the fountain height to 12 cm.  Following the
three-pulse interferometer sequence, the population distribution
of the atoms in each ensemble is measured.  In order to extract
the  gravitationally-induced phase shift, the phase of the final
interferometer pulse needs to be scanned.  The scanning is
accomplished digitally with the AWG by scanning the phase of the
RF waveform applied to the low frequency AOMs in the Raman beam
paths.  This phase scanning is in addition to the frequency chirp
applied during the pulse.  The total cycle time is 0.7 - 1.4 s,
depending on the trap loading parameters.

\subsection{Detection System}
\subsubsection{Detection Apparatus}
In order to realize high interferometer sensitivity, atoms must be detected with a high SNR following the interferometer pulse sequence. Our detection system uses a balanced, modulation transfer technique to reduce laser-induced detection noise and differentiate cold atoms from thermal background atoms.  This detection method is described in \cite{mcg00} and is summarized here.

Balanced detection uses two parallel, horizontally propagating probe beams, separated vertically by 2.5 cm, to detect atoms in one or both states following the interferometer. The absorption of the probe beams by the atoms is detected on a balanced photodiode with two inputs that are subtracted before amplification. Typically, the absorption by the cold atoms is around $0.1\%$.

We use two detection pulses as follows. Following the interferometer, atoms in the F = 4 state are detected by pulsing on near-resonant probe beams for 4.6 ms when the atom cloud is in the upper detection beam.  During this pulse, the mean velocity of the F = 4 atoms is slowed to rest by optical forces induced by the detection beams. Atoms in the F = 3 state continue to fall, however.  When these atoms intersect the lower probe region, atoms in the F = 4 state have not moved out of the upper detection region.  Pulsing on the probe beams for another 4.6 ms following a short repumper pulse measures the differential absorption between the two hyperfine states.  These two pulses provide sufficient
information to infer the overall transition probability in a way which is immune to shot-to-shot fluctuations in the total atom number.  We have shown that this method is immune to laser frequency and amplitude technical noise as well \cite{mcg00}.

In addition, a modulation transfer technique is employed to remove
noise from the thermal background vapor. Here a frequency
modulated (FM) pump beam is applied along a vertical axis
orthogonally to the probe beams during each detection pulse.  The
nonlinear interaction between the FM pump and the atoms modulates
the complex index of refraction of the atoms. This modulation
produces an amplitude modulation (AM) on the probe beams
\cite{shi82,syn80}.  The AM is detected by the balanced detector,
mixed down to DC with a double-balanced mixer, and integrated with
a Hewlett Packard 3458A multimeter.  The orthogonality of the pump
and probe beams provides a velocity selectivity that allows
rejection of signals from the fast-moving thermal background
atoms.

\subsubsection{Detection Noise Analysis}
The amplitude of the modulated absorption signal at the photodiode is about 0.8 pW per atom.  The detection photodiodes have noise-equivalent powers corresponding to a 60 atom detection sensitivity in the 4.6 ms measurement window.  The digital voltmeters that acquire the signal from the photodiodes are only slightly noisier with a noise floor corresponding to 100 atoms. The largest contribution to the intrinsic noise comes from the detection light itself. Because absorption detection is used and the absorption is small, there is a substantial amount of unabsorbed light, $> 99\%$, striking the detector.  Shot noise on the number of photons incident on the detector during the
integration window is the leading intrinsic noise source.  The shot noise power is 0.25 nW, resulting in a minimum detectable signal of $\sim 300$ atoms. This noise dominates the technical noise sources.

There is also a noise component of similar size due to a small
number of atoms in undesired states which survive the state
preparation.  Slight changes in laser frequency and selection
pulse efficiency during the state preparation cause this number to
fluctuate.  However, this noise source is common between the two
chambers and is also suppressed by the balanced detection method.

To summarize, the noise of our detection system allows detection
of transition probability at the atom shot-noise limit when we
detect more than $\sim 10^5$ atoms.  These noise limits are
discussed in the  context of our overall instrument sensitivity
below.

\subsection{Signal Extraction}
\subsubsection{Interference Fringe Fitting}
As described above, the phase of the final interferometer pulse is
scanned electronically.  A straightforward method of extracting
gravity gradient information is to determine the gravitationally
induced phase shifts in each atom ensemble by performing least
squares sinusoidal fits on the observed interference fringes. This
is possible when vibration-induced phase noise is $\lesssim 1$
rad.  The gravity gradient is obtained by subtracting the two
phase shifts from each other.  Vibrational phase noise and local
oscillator phase noise cause the phase extracted by the sine fits
to be shifted.  However, these noise sources couple to the two
accelerometers in a common-mode way. This common-mode behavior
results in the two sinusoidal fits being shifted by an identical
amount, and any effect of common-mode noise is cancelled in the
subtraction used to obtain the gravity gradient.  We study the
statistics of the resulting phase differences under static gravity
gradient conditions to estimate instrument noise.

We find that the distribution of the residual noise contains outlying points. Eliminating these outlying  points increases the SNR by up to a factor of six. The reduction of the number of points is incorporated into the data collection time in determining the  sensitivity.  We are presently studying possible sources of this noise component.

The ratio of the interference fringe amplitude to the the standard deviation of the phase difference distributions determines the instrument SNR.  The side of a fringe, {\it i.e.} the linear slope
of a sine wave, is most sensitive to phase shifts, with a sensitivity given by $\delta\phi$ = 1/SNR. For gravitationally induced phase shifts, the sensitivity to a change in the gravitational acceleration is $\delta$g $ = \delta\phi /(2$kT$^2$)t$^{1/2}$, where t is the data acquisition time to achieve the uncertainty $\delta\phi$. Dividing by the chamber separation determines the sensitivity to gradients.

\subsubsection{Magnetic Phase Shifting}
Good common-mode noise suppression requires that the lower and
upper chamber fringes be acquired in phase.  However, the Earth's
gravity gradient of $\sim$ 3000 E will cause a relative phase
shift of $\sim$ 1.5 rad between the two chambers.  In order to
accommodate this shift, a bias magnetic field is pulsed on for 67
ms in the lower  chamber during the atom interferometer.  This
field pulse causes a phase shift due to the second-order Zeeman
effect.  The amplitude of this pulse is chosen to produce a shift
which compensates the shift due to the gravity gradient, allowing
both fringes to be acquired in phase.

We have quantitatively studied the effectiveness of common-mode
noise suppression as a function of the phase alignment of the
interferometer fringes.  Fig.~\ref{fig3} shows a comparison of the
predicted dependence of the noise on the relative phase between
fringes with experimental SNR data.  For simplicity, we scanned
the relative phase between pulses simply by changing the
interrogation time T. (For this study, we disabled the magnetic
phase shifter and worked with relatively short interrogation
times.) The data shown in Fig.~\ref{fig3} is the SNR of the phase
difference between the two fringes obtained by performing least
squares fits as described above on a set of interference fringes.  The prediction is the simulated SNR of the phase difference between two phase-mismatched sine functions with phase noise added.  The predicted SNR is added in quadrature with uncorrelated detection noise noise in order to compare the simulation with experimental data in this curve.  As shown, the data is in good agreement with the theory, demonstrating the need for the magnetic phase shifting pulse.

The stability of the bias field used in the magnetic shifter has
also been studied.  Noise on this bias field could lead to extra
phase noise on the gravity gradient signals.  This noise has been
investigated by applying a bias pulse in a microwave clock $\pi/2$-$\pi/2$ experiment and comparing the SNR for a weak pulse
(phase shift $\sim$ 1.2 rad) and a strong field pulse (phase shift
$\sim$ 67 rad).  For the large phase shift,  no extra phase noise
or drift is seen at a fractional uncertainty in the phase of less
than $3\times10^{-6}$.  This implies that for a 1 rad shift, the
bias pulse contributes noise at a level below $8\times10^{-13}$ g.

\subsubsection{High Phase Noise Regimes}
In the case where phase noise is greater than 1 rad, the noise
renders it impossible to characterize instrument noise using the
least squares method described above.  In the most sensitive modes
of operation, the vibrations of the reference platform induce
phase noise much larger than this level.  A different analysis
technique must be performed in this regime, using a point-by-point
analysis. After collection of the gradiometer data, a signal
extraction algorithm removes amplitude and phase noise from the
signals: Our detection method allows measurement of the number of
atoms in the F = 4 state and the population difference between the
two states. With this information, common amplitude fluctuations
in each chamber, primarily from number fluctuations from the MOT
loading, can be removed.  After removing amplitude noise, a
cross-chamber normalization is performed to reject phase noise
that is primarily vibration-induced.

The details of this noise analysis procedure are as follows. After
each interferometer cycle (which represents one gradient measurement) two samples are acquired in each accelerometer:
signals proportional to the F = 4 population and to the
differential population ({\it i.e.} proportional to the number of
atoms in F = 4 minus the number of atoms in F = 3).  The two
samples are combined to infer the total number of atoms present in
the interferometer during each experimental cycle.  The F = 4
signal is then divided by this total atom number to remove any
fluctuations in the amplitude of the F = 4 signal from
shot-to-shot atom number fluctuations.  We normalize each
interferometer with this procedure. To remove common phase noise
between the two chambers, a series of experimental cycles is
taken, and a least squares minimization ({\it via} Gaussian
elimination) is performed on the quantity (S$_1 - \alpha $S$_2
-\beta)^2$ where S$_1$ and S$_2$ are the shot-by-shot normalized
F = 4 population levels from the two interferometers.  The fit
constants $\alpha$ and $\beta$ are used to compensate for possible
differences in interference contrast between the two
accelerometers.

The residuals of the Gaussian elimination procedure are used to estimate instrument noise, as show in Fig. ~\ref{fig4}.  Again, this distribution is non-Gaussian, and outlying points are discarded to obtain SNR and short-term sensitivity estimates.

Note that while this method is suitable for characterization of
the instrument noise floor under static gradient conditions, further work is need to demonstrate effective algorithms for extraction of dynamic gradient signals (such as would be present in  moving-platform applications).  We are currently exploring algorithms for this purpose.

\subsection{Vibration Isolation Subsystem}
At the most sensitive gradiometer operation, vibration phase noise is large, and the high phase noise algorithm is employed. However, in order to verify the validity of this algorithm, a vibration isolation system was constructed to remove most of the vibration-induced phase noise from the interference fringes.  With this reduction in phase noise, the least squares fit algorithms also can be used to reduce the data.

\subsubsection{Mechanical Design}
The primary object in the instrument that must be isolated from vibrations is the Raman beam corner cube.  All other optics are positioned so that any vibrations Doppler shift the two Raman beams in a common way, and the Raman difference frequency remains unchanged.  The Raman beam corner cube retroreflector is mounted on a Newport sub-Hertz platform (SHP) which provides the principal vibration isolation (see Fig.~\ref{fig5}).  The SHP is guided by a linear air bearing (New Way S4010002) along the vertical axis.  The SHP provides isolation in the range of 0.5 Hz - 40 Hz. An accelerometer (Teledyne Geotech S-510) is mounted on the SHP to monitor platform accelerations. The corner cube is attached to the platform by a stack of two pieces of 1 in.~thick lead filled acoustic foam separated by a 0.5 in.~thick sheet of aluminum.  The double stack of acoustic foam reduces vibrations of 30 Hz and higher by more than 20 dB. A linear voice coil actuator provides active feedback to the SHP.  Additionally, the voice coil can be used to drive platform accelerations; this shake-testing is the subject of Section III.B.

\subsubsection{DSP Servo System}
Here we describe the active servo system for the SHP platform. The active feedback loop starts with the accelerometer to monitor vibrations.  The accelerometer output is processed by a DSP (Spectrum Signal Indy TMS320C32), which we use to filter digitally the accelerometer input (as described below) and generate the feedback error signal.  The feedback signal, after being buffered by a voltage amplifier, closes the feedback loop by driving the voice coil mounted between the SHP and the platform support. We apply the following digital filters in processing the accelerometer signal. First, a lag filter with a bandwidth of 1 Hz to 80 Hz rolls off the feedback below the accelerometer's 100 Hz high frequency cut-off.  Next, a second lag filter with identical bandwidth is used to make the gain roll-off second order. Finally, two lead filters are applied to keep the system from oscillating at low frequency near the closed-loop SHP resonance of 0.03 Hz, which is also close to the internal highpass frequency of the
accelerometer.  The two lead filters have bandwidths of 38 mHz - 200 Hz and 380 mHz - 200 Hz respectively. The total gain of all four filters is 1600. This work is similar in concept to that reported in Ref. \cite{hen99}.

Using this servo, we are able to reduce the vibrations to near the noise floor of the accelerometer (10$^{-8}$ g/Hz$^{1/2}$) over a  bandwidth of 40 mHz - 25 Hz.  Higher frequencies are passively attenuated by the acoustic foam.  With the addition of the vibration isolation system, phase noise from accelerations of the corner cube is reduced to less the 1 rad, and least squares sinusoidal fits may be performed on the fringes for the longest interrogation times.

\subsection{Microwave Generation}
The generation and delivery of the 9.2 GHz microwave field is briefly described here.  The microwave field is coupled to the atoms through RF horns attached to viewports on the MOT chambers. The microwave frequency is tied to a 10 MHz reference, temperature stabilized, master crystal oscillator (Oscilloquartz OCXO, stability of 1.4$\times$10$^{-13}$ in 1 s).  The reference oscillator drives a 100 MHz phase locked oscillator (PLO, Wenzel 500-0732) which is the input to a Microlambda (MLPE 1162) 9.2 GHz PLO.  The RF is mixed in a single sideband mixer with an $\sim$ 7.4 MHz signal from an AWG which is also phase locked to the reference oscillator.  The AWG is used to scan the RF frequency and phase. The mixer output is amplified up to $\sim$ 1 W and sent to the horns.  The RF power is controlled by the AWG output, and the relative power to the two chambers can be adjusted with appropriate attenuation in the two paths.

We perform a microwave $\pi$/2 - $\pi$/2 clock experiment as a diagnostic to check the phase noise performance of our oscillators and the noise performance of our detection system. We have shown that we can detect microwave clock fringes with 1000:1 SNR using our normalized detection \cite{mcg00}.  This SNR is at the atom shot noise limit for our fountain with no velocity selection ($\sim$ 10$^6$ atoms/shot).

\section{Instrument Performance}

\subsection{Sensitivity}
Each interference fringe is typically recorded with 15 consecutive cycles of the experiment.  The number of points per scan is kept small to decrease sensitivity to long term drifts in signal amplitude and contrast.  The source of such drifts include fluctuations in laser power and drifts in the Cs vapor pressure in the vacuum chambers.  The normalized data reduction method described in Section II.F is performed on the data.  The resulting observed SNR is typically 150:1.  This corresponds to a differential accelerometer performance of $4\times10^{-8}$ m/s$^2$/Hz$^{1/2}$, or $\sim 4\times10^{-9}$ g/Hz$^{1/2}$. Scaling this to a 10 m baseline gives an inferred gradient sensitivity of 4 E/Hz$^{1/2}$. Fig.~\ref{fig4} shows typical high sensitivity fringes in the upper and lower chamber for a T = 157.5 ms
interferometer. The fringe contrast is typically $\approx$ 33\%.

Fig.~\ref{fig6} compares interference phase scans with the reference platform rigidly attached to the optical table ($\sim 10^{-5}$ g technical acceleration noise) with the reference platform servoed using the vibration isolation system.  We can use the data acquired under servoed conditions ({\it e.g.}  Fig.~\ref{fig6}b) to provide an independent check on the observed SNR, by performing the least-squares fitting analysis detailed in section II.F.1, and  comparing it with the point-by-point analysis of II.F.3 for data  acquired without the servo ({\it e.g.} Fig.~\ref{fig6}a).  We find that both approaches yield consistent noise estimates.

\subsubsection{Noise}
The dominant noise source is atom shot noise: the Poissonian fluctuations that arise from detecting atoms in coherent superposition states, also called quantum projection noise \cite{ita93}. Atom shot noise scales as the square root of the number of atoms detected.  As long as there is a sufficiently large number of atoms contributing to the interference fringes, this noise source will be the dominant noise source.

For the T = 157.5 ms gradiometer, the demonstrated SNR is predominantly limited by atom number shot noise.  The fringe amplitude corresponds to about $2.5\times10^5$ atoms.  At 33$\%$ fringe contrast, there is a mean offset of $7\times10^5$ atoms. Based on atom shot noise, these atom numbers put a limit of 300:1 on the SNR for each interference fringe ($2.5\times10^5 / (7\times10^5)^{1/2}$).  Additionally, normalizing each chamber reduces the SNR for each gravimeter to (300/2$^{1/2}$):1. Subtracting the two gravimeter signals to produce a gradient signal decreases the SNR by another factor of 2$^{1/2}$, fundamentally limiting the SNR for the full gravity gradiometer to 150:1.

The noise produced from photon shot noise on the unabsorbed portion of the probe detection light, corresponding to $\sim$300 atom minimum detectable signal, is approximately at the 800:1 level for each individual fringe, giving about a 400:1 limit after the normalization and subtraction of the two fringes.  This means that photon shot noise does not impose a substantial SNR limit for the current number of atoms, but it would become significant for reduced atom numbers.

\subsubsection{Mass Detection}
In order to demonstrate further the sensitivity of the gravity gradiometer, measurement of the gradient of a nearby object was performed.  Previously, the Earth's gravitational gradient was measured \cite{sna98}.  With the improved sensitivity, a measurement of the gravity gradient from small test masses has been made.  For this measurement, eight lead bricks ($\sim$ 12.5 kg each) were stacked symmetrically about the lower chamber, $\sim$ 0.2 m from the apex of the atomic fountain. The calculated acceleration from this configuration of mass should yield a $8.2\times 10^{-9}$ g signal. The measured acceleration is $8.1 \pm 2.1 \times 10^{-9}$ g which agrees well with the expected phase shift.

We are currently extending this work to measure the gradient of a well characterized mass.  Our goal is to measure G, the gravitational constant, at the part per thousand level.

\subsection{Immunity to Environmental Noise}
An important feature of the gradiometer is the ability to reject common mode accelerations in the two measurements. This is critical for a device which might be used on a moving platform. In order to demonstrate this capability, we performed a series of acceleration and tilt tests to characterize the instrument's sensitivity to platform vibration and tilt noise.  We characterize the effects of accelerations and tilts by measuring instrument SNR as a function of the amplitude and frequency of an external platform drive.

\subsubsection{Linear Acceleration}
We characterized immunity to linear accelerations by shaking the platform on which the retroreflecting corner cube is mounted.  As in earlier work, this corner cube plays the role of the acceleration reference for the acceleration measurements \cite{kas92,sna98}.  We drive the platform by applying sinusoidal drive currents to the voice coil coupled to the vibration isolation platform.  We monitor the resulting platform acceleration with the platform accelerometer (described in Section II.G.1). For this work, the high frequency lead-foam passive isolation between the platform and the corner cube was removed. This study was done with the interferometer operating in its most sensitive configuration (at T = 157.5 ms interrogation time). Fig.~\ref{fig7} summarizes the results of these tests. For frequencies in the range 1 to 100 Hz no significant degradation of the SNR was observed for drive amplitudes up to $2.5\times10^{-2}$ g. This corresponds to a common-mode rejection ratio of 140 dB. At amplitudes greater than $\sim$ 0.1 g, the accelerations are large enough that individual Raman pulses are Doppler-shifted out of the Raman resonance condition.  At this point, we observe poor interference fringe contrast.

\subsubsection{Rotation}
Tilts of the Raman interferometer beams are expected to degrade the sensitivity of the measurement.  In order to study the effects of small tilt displacements, we floated the optical table on which the gradiometer apparatus was mounted using commercial pneumatic legs (Newport I-2000).  We drove tilt motions with an appropriately placed voice coil, and characterized these motions using a tilt meter (Applied Geomechanics 755-1129; specified resolution of 1 $\mu$rad, 20 Hz bandwidth). The rotation vector associated with the tilt motion was in a plane perpendicular to the Raman propagation axis.  Fig.~\ref{fig8} shows observed SNR versus maximum tilt amplitude for different driving frequencies.

Both centrifugal and Coriolis forces lead to a degradation in instrument sensitivity.  A rotation $\Omega$ generates a centrifugal acceleration $R\ \Omega^2$, where $R$ is the distance of one accelerometer from the center of rotation.  For two accelerometers separated by distance $\delta R$, the differential acceleration is $\delta R\ \Omega^2$, while the acceleration
gradient is $\Omega^2$. For our operating parameters, we expect this mechanism to begin to degrade the SNR at $\Omega \sim 10^{-4}$ rad/s.  However, full loss in SNR is not expected until $\Omega \sim 10^{-3}$ rad/s for this mechanism, much higher than the rotation rates employed in this study. The Coriolis force leads to a loss in fringe contrast. In this case, the source transverse velocity spread of $\delta$v $\sim$3 cm/s leads to an interference phase spread of $\delta\phi \sim 2$k$_{eff}$ $\delta$v $\Omega$ T$^2$.  At $\Omega \sim 10^{-4}$, $\delta\phi \sim$ 1 rad, and we expect a full loss in contrast. This is in reasonable agreement with our observations.

\subsection{Accuracy Estimation}
To demonstrate accelerometer (hence gradiometer) accuracy, we
monitored accelerometer outputs over extended periods of time.  As
in \cite{peters99} we observed the daily fluctuations in the
measured acceleration due to gravity induced by the motion of the
Sun and Moon. Fig.~\ref{fig9} shows data from one accelerometer
taken over a period of several days compared to the Tamura 87 tide
model \cite{tides}. The only free parameter in this fit is an overall phase offset.  From this data set, we constrain the accelerometer phase offset to better than $10^{-9}$ g over the two day measurement cycle. Subsequent measurements, taken several weeks later, resulted in a similar determination of the fitted offset. From these measurements we infer an accelerometer accuracy of $<10^{-9}$ g over time periods of days.  For a 10 m baseline
gradient instrument, this corresponds to an inferred accuracy of
better than 1 E.  Further accuracy studies are underway.

\section{Discussion}

In this section we discuss the light-pulse method in the context
of other de Broglie wave interference methods based on laser-atom
interactions, as well as alternate geometries for gradient
measurements.

\subsection{Interferometer Comparisons}
Here we compare de Broglie-wave gravimeters based on multiple
pulse techniques using stimulated Raman transitions, Bragg
scattering, diffraction in the Raman-Nath regime, adiabatic
transfer, and the AC Josephson effect with Bose-Einstein
condensates (BEC).

We have recently demonstrated large-area, multiple Raman pulse
techniques that may enable higher sensitivity gravimeters and
gradiometers\cite{mcg00b}.  By using extra stimulated Raman
transition pulses of alternating propagation direction, a large
relative momentum can be transferred to the two interfering atomic
wavepackets, resulting in a large-area interferometer. The
sensitivity of this device to gravitational phase shifts increases
linearly with the relative momentum imparted to the wavepackets,
or the number of extra pulse applied. For instance, a gradiometer
with one extra set of pulses has a relative splitting of 6$\hbar$k
and a sensitivity to gravity gradients given by $\Delta\phi =
6{\bf k} \cdot \Delta{\bf g}$T$^2$.  In principle, each subsequent
Raman pulse should not affect the fringe contrast, except for a
small amount of spontaneous emission.  This process should be
extendable to a large number of extra pulses and a large relative
momentum.

This method might be extended using special sequences of composite
pulses (see Section II.C.2).  The benefits of using composite pulses
in this context would be an extremely large input velocity
acceptance for a fixed Rabi frequency and robust suppression of
spatial inhomogeneities in Rabi frequency (which arise due to the
finite size of the Raman laser beams).  For example, consider the
following sequence. A high efficiency composite microwave pulse
could be used for the initial $\pi$/2 pulse.  Next a composite
optical $\pi$ pulse gives the wavepackets 4$\hbar$k momentum
splitting, and after a long drift time two composite $\pi$ pulses
redirect the wavepackets back towards each other. Finally, a last
composite optical pulse and another composite microwave pulse
recombine the wavepackets to complete the interferometer.
Simulations show that this larger area interferometer has the
potential to have a very high fringe contrast (because of the
symmetric composite pulses).

Bragg scattering-based interferometers diffract atoms from
standing waves of laser light.  As in the light-pulse
interferometers, these interactions can be configured as atom
optic  beamsplitters and mirrors \cite{bragg}. The primary virtue
of Bragg scattering interferometers is that the atoms always
remain in the same internal state.  This reduces the sensitivity
to systematic phase shifts such as Zeeman shifts and AC Stark
shifts, since the wavepackets in each arm of the interferometer
experience symmetric phase shifts due to these effects. In
comparison, the Raman method requires the use of propagation
vector reversal techniques to gain immunity to these possible
systematics.  Furthermore, high-order Bragg processes can be used
to create large-area, high-sensitivity instruments. However,
high-order Bragg processes operate efficiently over a relatively
narrow range of initial atomic velocities (significantly less than
a photon recoil velocity). This severely constrains the fraction
of atoms which can contribute to the interference signal, and thus
the atom counting rate. On the other hand, BEC or atom laser
sources may eventually produce extremely bright atomic beams
(having excellent velocity collimation).  In this case, Bragg
processes may become a competitive choice for interferometer
sensors.

A gravimeter based on diffraction in the Raman-Nath regime has
been demonstrated \cite{nath}.  In this interferometer, short,
intense pulses of light are applied to the atomic ensemble. Due to
the shortness and intensity of these pulses, atoms are diffracted
into a large number of diffraction orders.  Two such Raman-Nath
pulses are used to construct the interferometer, and because of
the wide spread of wavepacket momenta, many different interfering
paths exist.  Some of these paths overlap and interfere after an
echo time, and the echo time determines the sensitivity to
gravitational phase shifts.  The Raman-Nath diffraction also
populates many higher lying momentum states that do not contribute
to the closed interferometer paths, which decreases the
interferometer contrast and places severe constraints on the SNR.
High sensitivity gravimeters, which require good SNR for their
operation, have yet to be demonstrated using this approach.

Adiabatic transfer recently has been used in a proof-of-principle
demonstration of a possible large-area interferometer
\cite{foote}. In this approach, atoms are put into coherent
superpositions of two states using a microwave pulse.  Momentum is
transferred to one state in this superposition by adiabatically
transferring it from the m$_f$ = 0 sublevel to the highest (or
lowest) lying m$_f$ level. Adiabatic transfer is used subsequently to manipulate and ultimately to overlap the wavepackets. A final microwave pulse is then used to interfere these wavepackets. This method can transfer up to 2m$_f \hbar$k momentum to one arm of the interferometer. The primary limitation to the utility of this method is that the adiabatic transfer populates magnetic field sensitive sublevels. This makes the interferometer sensitive to Zeeman shifts. While such an interferometer may demonstrate a high sensitivity, it remains to be seen whether it can achieve high  accuracy.

Finally, the AC Josephson effect in arrays of Bose-Einstein
condensed atoms \cite{anderson} has recently been used to make a
proof-of-principle gravity measurement. In this approach,
condensate atoms tunnel from an array of vertically spaced lattice
sites. Atoms tunnelling from different sites subsequently
interfere.  The resulting interference pattern is a periodic train
of atom pulses whose frequency depends on the strength of the
gravitational potential. This frequency can be measured with high
accuracy using conventional atom detection techniques.  A major
technological drawback to this technique is the need for
Bose-Einstein condensed atomic sources, which still are difficult
to produce and not suitable for a portable apparatus. Also, the
time required to condense atoms is typically in excess of 30 s,
which results in a low instrument bandwidth.  However, if
techniques to produce robust, BEC atom sources improve, this
method may become viable for future instruments.

\subsection{Direct Gradient Measurements}

\subsubsection{Multi-loop Interferometers}
There are several  Raman-pulse based schemes which can be used for
direct gravity gradient measurements.  The simplest of these is
the double-loop, or figure-eight, interferometer\cite{cla88}. This
geometry can be achieved in the light-pulse method by inserting an
extra $\pi$ pulse into the light-pulse sequence. Instead of
applying a $\pi/2$ pulse for the third pulse, the two atomic
wavepackets are allowed to pass through each other and form a
second loop [see Fig.~\ref{fig10}(b)].  Next a second $\pi$ pulse
redirects the wavepackets to close the second loop, and a final
$\pi/2$ pulse interferes the wavepackets.  This pulse sequence
directly produces a phase shift proportional to the the gravity
gradient by essentially performing a coherent subtraction of two
spatially, and temporally, separated gravity measurements (one for
each loop). Following the pulse rules presented in Section I, the
double-loop phase shift is given by $\Delta\phi =
\phi(t_1)-2\phi(t_2)+2\phi(t_3)-\phi(t_4)$, where $\phi(t_i)$ is
the phase of the i$^{th}$ Raman pulse at the position of the
atomic wavepacket at time t$_i$.  For a given fountain height the
double loop interferometer is maximally sensitive when the
interferometer spans one half of the total fountain time.
Evaluating this phase, assuming a linear gravitational gradient
g(z) = g$_\circ$ + $\alpha$z along the sensitive axis, gives
$\Delta\phi \approx 8$kg$_{\circ}\alpha$T$^4$, where g$_\circ$ is
the gravitational component along the Raman wavevector at the
atoms' initial position and $\alpha$ is the linear gradient.  This
formula is valid for an interferometer in the first half of a
fountain of length 8T and for ${\alpha}$z$^\prime$ $\ll$ g$_\circ$
where z$^\prime$ is the total height of the fountain.

One problem with this approach is that if the interferometer spans
the whole fountain time, {\it i.e.} is symmetric about the
fountain's peak, then the gravity signal from each loop will be
identical, resulting in no phase shift.  Thus, to obtain gradient
sensitivity, the double-loop must be used only in the first half
or the second half of the fountain, which severely constrains the
possible instrument sensitivity.  (This means the maximum
interaction time for our fountain is T = 39 ms, resulting in 3000
times less sensitivity to gradients as compared with two single
loop interferometer which are separated by 10 m.) In addition to
this sensitivity limit, it is difficult to make the gradiometer
baseline arbitrarily large and there is no common-mode vibration
rejection, since each of the two acceleration measurements which
comprise the gravity gradient signal are made at differing times.

A slight modification of the double-loop method allows full use of
the fountain interaction time, resulting in a three-fold increase in sensitivity from a double-loop interferometer in an apparatus of equivalent size.  This modified sequence uses an extra $\pi$ pulse in a $\pi/2-\pi-\pi-\pi-\pi/2$ sequence, which creates a triple-loop interferometer as seen in Fig.~\ref{fig10}(c).  The phase shift is $\Delta\phi = \phi(t_1)-2\phi(t_2)+2\phi(t_3)-2\phi(t_4)+\phi(t_5)$, which reduces to: $\Delta\phi \approx [17/3+4\sqrt{2}]$kg$_{\circ}\alpha$T$^4$, again assuming a linear
gradient and an interferometer that now spans the entire fountain
time, $(8+4\sqrt{2})$T.  This interferometer may be symmetrically
spaced about the fountain's apex, so that the maximal interrogation time is T = 46 ms for our fountain.  Interference fringes from a triple loop interferometer are shown in Fig.~\ref{fig11}.  The 10 m single-loop, two chamber gradiometer still is 1000 times more sensitive due to its larger baseline and longer effective interrogation time.  In addition, the triple-loop gradiometer, like the double-loop, does not provide vibration rejection, has a limited baseline, and is only slightly less sensitive to magnetic fields.

\subsubsection{Curvature Measurements}
Two multiple-loop gradiometers may be used in conjunction to measure the second moment of the gravity field, in a configuration similar to that used to measure the gradient with two single loop interferometers, as illustrated in Fig.~\ref{fig2}.  In this case, the triple-loop gradient phase shifts obtained from two spatially separated atomic ensembles are subtracted to obtain the second order curvature of the gravitational field.  This device does have immunity to spurious vibrational noise because measurements are made simultaneously with respect to a common platform. Proof-of-principle data is shown in Fig.~\ref{fig11}. Measurement of the second moment of the gravitational field allows differentiation between massive, distant objects and less massive, nearby objects that our gradiometer could not distinguish.

A more direct and sensitive way to characterize the second order
moment is to operate simultaneously three single-loop accelerometers spaced equidistantly along a single axis.  The difference between gradients obtained by differencing the first and second accelerometer outputs from the second to third accelerometer outputs gives a second moment measurement.  This device should have the same benefits as the previous device and would be significantly more sensitive.

\section{Conclusion}
Future sensitivity enhancements are likely.  Improvements in the
number of atoms trapped and optically prepared in the correct
state would allow for higher, atom-shot noise limited  SNRs. This
might be accomplished using recently demonstrated atom trapping
and cooling techniques \cite{jessen,treut}. Multiple pulse ($>$
2$\hbar$k) techniques will provide straightforward access to
large-area, high sensitivity configurations.

It should be noted that the performance of a gravity gradiometer
in a microgravity environment would be greatly enhanced due to the
larger available interrogation time ($\Delta\phi \sim T^2$)
without need for an atomic fountain, and such a high sensitivity
device would be ideal for tests of fundamental theories.

Finally, the current sensitivity makes measurement of
geophysical gradient signals possible.  A portable absolute
gradiometer would be useful for navigation, geodesy, terrain
estimation, and oil and mineral exploration.

{\bf Acknowledgements.}  We thank Todd Gustavson and Arnauld
Landragin for constructing the trapping laser lock and Kurt
Gibble for useful discussions regarding trapping and optical
pumping.

This work was supported by grants from the ONR, NASA, MURI, and
NSF.

\begin{figure}
\caption{A comparison of the transfer efficiency of a composite $\pi$ pulse (solid traces) with a regular $\pi$ pulse (dotted traces).  The detuning is from the F = 4 m$_f$ = 0 $\rightarrow$ F = 3 m$_f$ = 0 transition.}
\label{fig1}
\end{figure}

\begin{figure}
\caption{Schematic illustration of the
detection apparatus showing the Raman beam racetrack setup.}
\label{fig2}
\end{figure}

\begin{figure}
\caption{The decrease in the gradiometer SNR is shown due to a phase mismatch induced by the Earth's gravity gradient.  The phase shift $\Delta\phi$ is the gradient phase shift from increasing the interferometer time T ($\Delta\phi \propto $T$^2$).  The solid line is a theory based on the noise from two mismatched sine waves
in quadrature with photon shot noise from the detection system.}
\label{fig3}
\end{figure}

\begin{figure}
\caption{Gradiometer interference fringes with T = 157.5 ms.
Squares and circles represent the normalized upper and lower
chamber fringes respectively, after compensating for contrast
differences between the chambers. Triangles show the lower chamber
signal residuals.}
\label{fig4}
\end{figure}

\begin{figure}
\caption{Schematic of the vibration isolation apparatus featuring
elements for passive and active isolation.} \label{fig5}
\end{figure}

\begin{figure}
\caption{Gradiometer interference fringes with T = 157.5 ms.
Circles represent the raw data points, and the solid lines is a
sinusoidal least squares fit.  (a) shows a fringe with the
inertial reference platform, {\it i.e.} the cornercube, rigidly
attached to the optical table.  (b) shows an interference fringe
taken with the cornercube mounted on the vibration control
system.} \label{fig6}
\end{figure}

\begin{figure}
\caption{Results of the reference platform shake test.  No SNR
reduction is seen when driving accelerations on the reference
platform at amplitudes of $10^{-2}$ g over the frequency band
indicated.  This amounts to 140 dB of vibration rejection.  The observed maximum SNR is slightly reduced from the data of Fig.~\ref{fig5}}
\label{fig7}
\end{figure}

\begin{figure}
\caption{Results of the platform tilt test.  The floated optical
table is tilted at a number of frequencies, and the gradiometer is
insensitive to a range of tilt amplitudes.   The observed maximum SNR is slightly reduced from the data of Fig.~\ref{fig5}} \label{fig8}
\end{figure}

\begin{figure}
\caption{Gravitational tidal signals as monitored by one
accelerometer output over two days.  Data are scattered points and
the solid line is the tidal model with no free scaling parameters.} \label{fig9}
\end{figure}

\begin{figure}
\caption{Recoil diagrams for various interferometer pulse sequences.  Dotted lines represent the F = 4 states and solid lines are the F = 3 state. The sensitivity to accelerations is proportional to the area enclosed in the recoil diagrams.  The timing of $\pi$ and $\pi$/2 pulses are shown with the vertical arrows. a) Single-loop accelerometer.  b) Double-loop gradiometer.  c) Triple-loop gradiometer.} \label{fig10}
\end{figure}

\begin{figure}
\caption{Typical interference fringes from T = 44 ms triple loop
interferometers in the upper and lower chamber.  The solid lines
are least squares fits.  The phase difference between the two
fringes represents a measurement of the second moment of the
gravitational field.} \label{fig11}
\end{figure}


\begin{thebibliography}{99}

\bibitem{dam94} T. Damour and A. Polyakov, Nucl. Phys. B {\bf 423}, 532 (1994).

\bibitem{ven92} B. Venema, P. Majumder, S. Lamoreaux, B. Heckel and E. Fortson, Phys. Rev. Lett. {\bf 68}, 135 (1992).

\bibitem{moo00} P. Moore and B. Taylor, Rev. Mod. Phys. {\bf 72}, 351 (2000).

\bibitem{sna98} M.J. Snadden, J.M. McGuirk, P. Bouyer, K.G. Haritos and M.A. Kasevich, Phys. Rev. Lett. {\bf 81}, 971 (1998).

\bibitem{ber97} {\it Atom Interferometry}, edited by P. Berman (Academic Press, New York, 1997).

\bibitem{peters99} A. Peters, K.Y. Chung, and S. Chu, Nature {\bf 400}, 849 (1999).

\bibitem{kas92} M. Kasevich and S. Chu, Appl. Phys. B. {\bf 54}, 321 (1992).

\bibitem{kas91b} M. Kasevich and S. Chu, Phys. Rev. Lett. {\bf 67}, 181 (1991).

\bibitem{jek93} C. Jekeli, Geophysics {\bf 58}, 508 (1993).

\bibitem{bhp} See, for example, E. van Leeuwen, Leading Edge {\bf 19}, 1296 (2000).

\bibitem{rom} A.J. Romaides, J.C. Battis, R.W. Sands, A. Zorn, D.O. Benson, Jr., and D.J. DiFrancesco, Journ. of Phys. D: Applied Phys. {\bf 34}, 433 (2001).

\bibitem{moo93} M. Moody and H. Paik, Phys. Rev. Lett. {\bf 70}, 1195 (1993).

\bibitem{van94} F. van Kann, M. Buckingham, C. Edwards and R. Matthews, Physica B {\bf 194}, 61 (1994).

\bibitem{goo99} J. Goodkind, Rev. Sci. Instr. {\bf 70}, 4131 (1999).

\bibitem{bro00} J. Brown, T. Niebauer, F. Klopping and A. Herring, Geophys. Res. Lett. {\bf 27}, 33 (2000).

\bibitem{levitt86} M. H. Levitt, Progress in NMR Spectroscopy {\bf 18}, 61 (1986).

\bibitem{abragam61} A. Abragam, {\it the Principles of Nuclear Magnetism} (Oxford Univ. Press, Oxford, England, 1961).

\bibitem{bou96} P. Bouyer, T.L. Gustavson, K.G. Haritos, M.A. Kasevich, Optics Letters {\bf 21}, 1502 (1996).

\bibitem{mcg00} J.M. McGuirk, G.T. Foster, J.B. Fixler and M.A. Kasevich, Optics Lett. {\bf 26}, 364 (2001).

\bibitem{shi82} J.S. Shirley, Opt. Lett. {\bf 7}, 537 (1982).

\bibitem{syn80} J.J. Synder, R.K. Raj, D. Bloch and M. Ducloy, Opt. Lett. {\bf 5}, 163 (1980).

\bibitem{hen99} J.M. Hensley, A. Peters, and S. Chu, Rev. Sci. Tech. {\bf 70}, 2735 (1999).

\bibitem{ita93} W.M. Itano, J.C. Bollinger, J.M. Gilligan, D.J. Heinzen, F.L. Moore, M.G. Raizen, D.J. Wineland, Phys. Rev. A {\bf 47}, 3554 (1993).

\bibitem{tides} Y. Tamura, Bulletin d'Informations Mareés Terrestres {\bf 99}, 6813 (1987).

\bibitem{mcg00b} J.M. McGuirk, M.J. Snadden, and M.A. Kasevich, Phys. Rev. Lett. {\bf 85}, 4498 (2000).

\bibitem{bragg} D.M. Giltner, R.W. McGowan, and S.A. Lee, Phys. Rev. Lett. {\bf 75}, 2638 (1995).

\bibitem{nath}S.B. Cahn {\it et al.}, Phys. Rev. Lett. {\bf 79}, 784 (1997).

\bibitem{anderson} B.P. Anderson and M.A. Kasevich, Science {\bf  282}, 1686 (1998)).

\bibitem{foote} P.D. Featonby {\it et al.}, Phys. Rev. Lett. {\bf 81}, 495 (1998).

\bibitem{cla88} J. Clauser, Physica B {\bf 151}, 262 (1988).

\bibitem{jessen} S.E. Hamann, D.L. Haycock, G. Klose, P.H. Pax, I.H. Deutsch, and P.S. Jessen, Phys. Rev. A {\bf 61}, 1972 (1998).

\bibitem{treut} P. Treutlein, K.Y. Chung, and S. Chu, Phys. Rev. A {\bf 63}, (2001).

\end{thebibliography}
\end{document}